# Analyzing the Impact of Automated User Assistance Systems: A Systematic Review

MURAT ACAR, Bogazici University, Istanbul, Turkey
BEDIR TEKINERDOGAN, Wageningen University, Wageningen, The Netherlands

*Context*: User assistance is generally defined as the guided assistance to a user of a software system in order to help accomplish tasks and enhance user experience. Automated user assistance systems are equipped with online help system that provides information to the user in an electronic format and which can be opened directly in the application. Various different automated user assistance approaches have been proposed in the literature. However, there has been no attempt to systematically review and report the impact of automated user assistance systems.
*Objective*: The overall objective of this systematic review is to identify the state of art in automated user assistance systems, and describe the reported evidence for automated user assistance.
*Method*: A systematic literature review is conducted by a multiphase study selection process using the published literature since 2002.
*Results*: We reviewed 575 papers that are discovered using a well-planned review protocol, and 31 of them were assessed as primary studies related to our research questions.
*Conclusions*: Our study shows that user assistance systems can provide important benefits for the user but still more research is required in this domain.

## 1. INTRODUCTION

User assistance is generally defined as the guided assistance to a user of a software system in order to help accomplish tasks and enhance user experience. The definition implies all forms of help available to a user. The traditional form of user assistance is a user manual in paper form which is separate from the system. The main obstacle of manual user assistance is the inherently reactive property (M. S. Ali, Babar, Chen, & Stol, 2010; S. Ali, Briand, Hemmati, & Panesar-Walawege, 2010; Alvarez-Cortes, Zayas-Perez, Zarate-Silva, & Ramirez Uresti, 2007; Ames, 2001)(Delisle & Moulin, 2002). This means that users only consult the documentation when they do not know how to proceed. The result is that they stop what they are doing, open the documentation, find the information they are looking for and then return back to the application. Because of this separate effort and disruption of the user's flow of work, users are very often reluctant to using help. The term *automated* user assistance (AUA) is considerably broad and mainly refers to the existence of user assistance systems that are working together with the pertaining system (Sondheimer & Relles, 1982). AUA systems are usually equipped with online help system that provides information to the user in an electronic format and which can be opened directly in the application. However, AUA is also broader than online help, and can also include procedural and tutorial information. In general, an important goal of AUA is also to enhance user experience, which defines the way a person feels about using a system.

It appears that AUA has been applied to different domains including education systems, gaming, defense, etc. In addition, various different AUA approaches have been proposed in the literature. In this context it would be worthwhile to assess the cost and benefit of AUA systems and identify the approaches that are useful and effective. Unfortunately, there has been no attempt to systematically review and report the impact of AUA systems. The previous studies like (Andrade, Paso, & Novick, 2008), (Alvarez-Cortes et al., 2007) and (Sondheimer & Relles, 1982) aim to provide a survey of current trends in this area, but their scope and objectives are too narrow to be taken as a roadmap. Besides, their aim is not to undertake the whole field.

This article reports on a systematic literature review (SLR) (Afzal, Torkar, & Feldt, 2009;

Babar & Zhang, 2009)(Harrison et al., 1999)(B. A. Kitchenham, Dyba, & Jorgensen, 2004; B. Kitchenham & Charters, 2007; B. Kitchenham et al., 2009) on AUA systems to provide evidence-based insights that can help both researchers and practitioners to gain a better understanding of automated user assistance together with the kinds of evidence provided to support those claims. Practitioners who are interested in applying AUA can use the SLR as a roadmap for finding and analyzing the relevant approaches and decide about their applicability. For researchers this SLR provides an overview of the reported automated user assistance approaches together with the strength of the empirical evidence of the identified approaches. Likewise, the SLR reveals those areas of AUA that are not addressed by the reported research or that require further research. In parallel, the SLR also points out the limitations of the current practice of AUA. Finally, the information extraction scheme we used to characterize the study context and study findings can be used to guide the activities of designing and reporting future empirical studies of AUA.

The systematic review is conducted by a multiphase study selection process using the published literature since 2002. We reviewed 575 papers that are discovered using a well-planned review protocol, and 31 of them were assessed as primary studies related to our research questions. Our study shows that user assistance systems can provide important benefits for the user but more research is required in this domain.

The remainder of the paper is organized as follows. Section 2 provides a short background of AUA systems. Section 3 describes the SLR method used in this study. Section 4 presents the results of the SLR. Section 5 presents the discussion and finally section 6 concludes the paper.

## 2. AUTOMATED USER ASSISTANCE SYSTEMS

User assistance can be realized in different ways. An analysis of the literature on user assistance shows that we can distinguish between different types of user assistance as depicted in Fig.1. We can distinguish among a) *external manual offline help systems* that provide manual usually paper-based help documentation b) *external automated user assistance* that provide electronic help that can usually be triggered within the application, and c) *embedded user assistance* that provides online embedded help d) *context-sensitive embedded user assistance,* that is embedded and that provides assistance based on the specific context parameters.

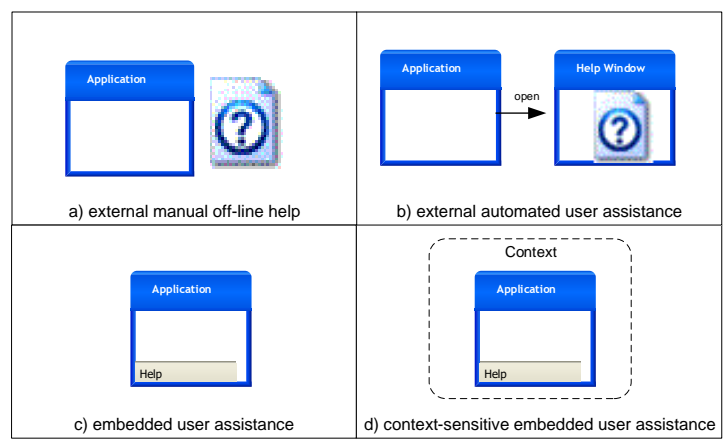

Fig.1. Four different categories of user assistance

For the categories a) and b) in which user assistance is externally defined the problems are almost similar. Users focus their attention on the completion of tasks in hand, and generally will be reluctant to using guides or any help instruments provided because of the interruption of the flow of work and the additional effort required. As such the trend is towards developing tools that provided embedded user assistance as defined in category c) and d). Embedded user assistance can be defined as documentation of the application that resides within the application. Embedded user assistance aim to minimize the disruption of the flow of work and relieve the effort of searching the required documentation by proactively delivering the information to users need when and where they need it. A special case of embedded user assistance is the development of *context sensitive user assistance* as defined in category d of Fig.1. In context-sensitive user assistance help is obtained from a specific point in the state of the software, providing help for the situation that is associated with that state. Opposite to general online user assistance, context-sensitive assistance does not need to be accessible for reading as a whole. In general the system is defined as a set of states to which a topic is related that extensively describes the corresponding state, situation, or feature of the software. In this paper and the SLR we focus on automated user assistance that includes categories b, c and d. We do not focus on approaches that target manual user assistance.

## 3. RESEARCH METHOD

A *systematic literature review* (SLR) or systematic review for short is a well-defined and rigorous method to identify, evaluate and interpret all relevant studies regarding a particular research question, topic area or phenomenon of interest (B. A. Kitchenham et al., 2004; B. Kitchenham & Charters, 2007; B. Kitchenham et al., 2009)(Mian, Conte, Natali, Biolchini, & Travassos, 2007). We conducted the SLR to reveal existing evidence concerning the automated user assistance systems. For this, we followed the complete guidelines for performing SLRs as proposed by Kitchenham and Charters (B. Kitchenham & Charters, 2007). In the following subsections we discuss our adopted research method that is based on an extensive review protocol.

### 3.1 Review Protocol

Before actually conducting the review we first defined the review protocol. A review protocol describes the methods that will be used to carry out a specific systematic review. Firstly, we specified our research questions based on the objectives of this systematic review. After this step we defined the search scope and the search strategy. The search scope defines the time span and the venues that we looked at. Once the search strategy was defined, we specified the study selection criteria (section IIID) that are used to determine which studies are included in, or excluded from, the systematic review. The selection criteria were piloted on a number of primary studies. We screened the primary studies at all phases on the basis of inclusion and exclusion criteria. Also, peer reviews were performed by the authors throughout the study selection process. Once the final set of preliminary studies was defined the data extraction strategy was developed which defines how the information required from each study is obtained (section IIIF). For this we developed a data extraction form that was defined after a pilot study. In the final step the data synthesis process takes place in which we present the extracted data and associated results.

### 3.2 Research Questions

As was previously stated, there has been no systematic review on automated user assistance systems. Hence, we aimed at the current state of art by means of our research questions with a wide sphere of influence in order to get the best evidence. The research questions were formulated as follows:
-   RQ 1. In which domains have automated user assistance techniques been applied?

- RQ 2. What are the existing research directions within automated user assistance?

**3.3 Search Strategy**

To answer the research questions as defined in the previous section we have conducted an extensive search of papers. In the following we describe the scope of the search, the adopted method and the search string.

*1) Scope*

Our search scope included the papers that were published starting from January 2002. The main motivation for 2002 was that user assistance conferences started just after this date. We searched for papers in selected venues publishing papers on user assistance and venues that publish high quality papers. We used the following search databases: IEEE Xplore, ACM Digital Library, Wiley Inter Science Journal Finder, ScienceDirect, Springer Link, ISI Web of Knowledge, and other channels including Google search and other web search engines. These venues are listed in Table 1. Our targeted search items were journal papers, conference papers, workshop papers, and white papers.

*2) Search Method*

To search the selected databases we used both manual and automatic search. Automatic search is realized through entering search strings on the search engines of the electronic data source. Manual search is realized through manually browsing the conferences, journals or other important sources.

TABLE 1: PUBLICATION SOURCES SEARCHED

| Source | NUMBER OF INCLUDED STUDIES AFTER APPLYING SEARCH QUERY | NUMBER OF INCLUDED STUDIES AFTER EXCLUSION CRITERION 1 | NUMBER OF INCLUDED STUDIES AFTER EXCLUSION CRITERION 2 |
|---|---|---|---|
| IEEE Xplore | 271 | 23 | 9 |
| ACM Digital Library | 153 | 21 | 6 |
| Wiley Interscience | 18 | 3 | 1 |
| Science Direct | 41 | 12 | 4 |
| Springer | 36 | 18 | 5 |
| ISI Web of Knowledge | 30 | 6 | 4 |
| Other Channels | 28 | 12 | 3 |
| Total | 575 | 93 | 32 |

*3) Search String*

Since we decomposed our research questions into some distinct facets (i.e. *population* and *intervention*), the designation of a search string could be accomplished according to the words determined in these facets (Dieste & Padua, 2007). Also, a list of synonyms, abbreviations, and alternative spellings was composed as an auxiliary instrument. Hereafter, a multifaceted search string could be obtained by means of Boolean ANDs and ORs. The following represents the search String that we defined for IEEE.

("user" AND "assistance") AND
("context sensitive" OR "context-sensitive" OR " process-sensitive" OR "process sensitive" OR "context aware" OR "context-aware" OR "embedded" OR "intelligent" OR "adaptive")

The search strings for other venues are listed in Appendix-III. Although the structure of search strings seems to be different, they are semantically equivalent. As was previously stated, in addition to the database searches, we also conducted manual searches both as a preliminary analysis and as a subsequent analysis after having observed the publication channels returned by the search strings. The manual searches appeared to be quite useful since we retrieved some good-quality articles that an automatic search could not reveal.

The result of the overall search process after applying the search queries and the manual search is shown in the second column of Table 1. As it can be seen from the table we could identify 575 papers at this stage of the search process.

**3.4 Study Selection Criteria**

In accordance with the SLR guidelines (B. Kitchenham & Charters, 2007) we further applied two exclusion criteria on the large-sized sample of papers in the first stage. The overall exclusion criteria that we used were as follows:

Exclusion criteria 1:
- Do not relate to a specific field of computer science
- Do not relate to user assistance
- Do not state any application of techniques, algorithms or methods to provide user assistance
- Do not report any results on the earnings of the approach proposed

Exclusion criteria 2:
- Abstracts or titles that do not mainly discuss the provision of user assistance were excluded
- Abstracts or titles that do not propose an approach to automate user assistance on the basis of the alternate terms that we have discussed were excluded

The exclusion criteria were checked by a manual analysis. After applying the first exclusion criteria 93 papers of the 575 papers remained, and finally after applying the second exclusion criteria 31 primary studies were selected.

**3.5 Data Extraction**

We recorded the places where the extracted information existed within the primary studies in spreadsheets. In order to support the process of synthesizing the extracted data, the form in Table 2 was developed in a progressive way so that the transition was performed seamlessly.

TABLE 2: RESEARCH QUESTIONS AND DATA EXTRACTED

| Research Questions | | Data Extracted |
|---|---|---|
| RQ1 | | main theme of the study, motivation for the main theme, targeted domain, publication details |
| RQ2 | RQ2.1 | study aims, automated user assistance solution used, research method used, examples of application of solution |
| | RQ2.2 | constraints/limitations, implications for future research and practical use, findings, major conclusions |
| RQ3 | | assessment approach |

**3.6 Data Synthesis**

Data synthesis is the process of collating and summarizing the extracted data in a manner suitable for answering the questions that an SLR seeks to answer. We made use of tabular representation of the data when feasible, and it enabled us to make comparisons across studies. Also, using the quantitative summaries of the results, we inferred the implications for future search, and consequently the existing research directions within automated user assistance.

### 4. RESULTS
### 4.1 Overview of Selected Studies

An overview of the primary studies according to publication channel is shown in Table 3. The table shows the publication channels, the types of articles and the number of studies that fall into the channels accordingly. From the table we can observe that the primary studies are identified from a diverse range of venues. One of the noteworthy publication channels is the "International Conference on Computer supported Cooperative Work in Design" in which the topics of agents and multi-agent systems, ontology and knowledge management, and collaborative design and manufacturing environments are contained. Also, the publication channel "ACM International Conference on Design of Communication (SIGDOC)" contained 3 studies in which user-centered design, methods, methodologies, and approaches are discussed.

TABLE 3: DISTRIBUTION OF STUDIES IN TERMS OF PUBLICATION CHANNEL AND OCCURRENCE

| Publication channel | Type | Number of studies |
|---|---|---|
| Computer Supported Cooperative Work in Design | Conference | 3 |
| SIGDOC | Conference | 3 |
| JASIST | Journal | 2 |
| Artificial Intelligence | Conference | 1 |
| Cooperative Information Agents | Conference | 1 |
| IEEE Transactions on Systems, Man, and Cybernetics | Journal | 1 |
| Web Intelligence | Conference | 1 |
| Cognitive Systems Research | Journal | 1 |
| Human - Centered Computing | Journal | 1 |
| Theoretical Issues in Ergonomics Science | Journal | 1 |
| Cognitive Ergonomics | Conference | 1 |
| Electronics, Robotics and Automotive Mechanics | Conference | 1 |
| Knowledge-Based Systems | Journal | 1 |
| New Generation Computing | Journal | 1 |
| Intelligent User Interfaces | Conference | 1 |
| Information Technology and Applications (ICITA) | Conference | 1 |
| Computers & Education | Journal | 1 |
| World Wide Web | Journal | 1 |
| Simulation Conference (WSC) | Conference | 1 |
| Dissertation | Thesis | 1 |
| Computers in Industry | Journal | 1 |
| IEEE Transactions on Visualization and Computer Graphics | Journal | 1 |
| Human-Computer Interaction | Journal | 1 |
| Information Processing and Management | Journal | 1 |
| Design, User Experience, and Usability | Journal | 1 |
| Transactions on Aspect-Oriented Software Development | Journal | 1 |

### 4.2 Research methods

Table 4 provides the list of research methods used in the selected 31 primary studies. There are five types of research methods that we looked for in the review. The numbers in the table reveal that most of the primary studies are based on either a single case study or an experiment. The reviewed survey-like study C made contributions also in the interpretation of qualitative primary studies. The study X both establishes a comparison reference between three different approaches and reviews the current trends and research efforts.

TABLE 4: STUDIES BY RESEARCH METHODS

| Research method | Studies | Number | Percent |
| --- | --- | --- | --- |
| Single-case | D, D, F, I, M, O, P, S, U, Y, Z, AA, BB, DD | 15 | 48.4 % |
| Multiple-case | T, EE | 2 | 6.5 % |
| Survey | C | 1 | 3.2 % |
| Experiment | A, B, F, H, H, K, K, L, R, V, W, CC | 12 | 38.7 % |
| Benchmarking | X | 1 | 3.2 % |

## 4.3 Systems Investigated

This section outlines the results we extracted related to three main research questions. We present the data extracted from the primary studies in the form of findings, separately for each research question.

**RQ1. In which domains have automated user assistance techniques been applied?**

Table 5 shows the categories of the target domains that we discovered. The categorization was done by descriptive and qualitative synthesis as proposed in (B. Kitchenham & Charters, 2007). As shown in Table 6 the category *Education* includes three subcategories including distance learning, remote experimentation and task experience. Study M discusses the user assistance for teachers in collaborative e-learning environments. Study AA and DD propose intelligent and adaptive user-assistance in learning environments for remote experimentation. Study D aims to enhance task experience for using excel sheets.

The domain of *Control Systems* defines the systems that deal with systems whose failure or malfunction may result in death or serious injury to people, or loss or severe damage to equipment or environmental harm. We can find here three studies (A, BB, EE) that have applied user assistance in sub-categories of air-traffic control, process control in power plants, wireless sensor networks, defense systems and smart environments.

TABLE 5: IDENTIFIED APPLICATION AREAS OF USER ASSISTANCE SOLUTIONS

| Domain | Identified Sub-Categories | Studies |
| --- | --- | --- |
| Education | Distance learning | M |
| | Remote Experimentation | AA, DD |
| | Task Experience | B, D |
| Control Systems | Air traffic control | A |
| | Process control in power plants | A |
| | Wireless sensor networks | A |
| | Defense Systems | EE |
| | Smart Environments | BB |
| Manufacturing | Manual Tasks | U |
| | Human-Machine Interaction | N |
| Collaborative Design | Interoperation of heterogeneous software agents | O, Q |
| | Collaborative design environments | M, Q, T, W, Y, Z |
| Healthcare | Human stress monitoring | B |
| | Emergency vehicle dispatchers | A |
| | Assistance to people with disability | A |
| Mobile Application | Mobile devices (smartphones) | G, V |
| | Ubiquitous computing | S |
| Web | Web Browsing/Navigation | F, J, K, M |
| | E-Commerce | J |
| | Information retrieval systems | I, K, L |
| User Interface | Graphical user interface | C, E, H, R, W, X |
| | User actions | P |
| | Haptic Guidance | CC |

In the domain of *Manufacturing* user assistance has been applied to supporting manual tasks and the access to machines by humans. Study U provides an approach for providing user assistance for manual tasks. Study N describes approaches for access to machine functions.

The domain of *Collaborative Design* defines systems in which different persons aim to produce a design collaboratively. As stated under the category Education, study M provides a user assistance approach for teachers in collaborative e-learning environments. Study T presents an agent-based approach, and study Y provides an application of user assistance in collaborative environments. Study O defines an approach for development of personalized user agents. Study Q describes an agent-based approach for personal assistance in collaborative design environments. Study Z provides a general approach for intelligent user assistance in collaborative design environments.

User assistance in healthcare has been applied to human stress monitoring, emergency vehicle dispatchers, and assistance to disabled people. Study A provides a dynamic probabilistic framework based on the dynamic Bayesian networks (DBNs) to dynamically model and recognize user's affective states and to provide the appropriate assistance in order to keep user in a productive state. Study B presents a decision theoretic model for stress recognition and user assistance.

In the domain *Web* the studies focused on web browsing, e-commerce, and IR systems. We could identify six different primary studies here.

The final category *User Interface* classifies the primary studies that focus on user interface concerns. The studies C,E,H,R,W,X,P and CC have been identified as the primary studies in this category.

Based on Table 5 several observations can be made. As we can see in the table, user assistance has been applied to a variety of domains and no single domain is dominant in this perspective. Also most of the primary studies address a single domain. The studies A and M address multiple domains.

**RQ 2. What are the existing research directions within automated user assistance?**
After the identification of the domains in which user assistance has been applied we aimed to identify the research directions in the state-of-the-art. As described in section IIIB this research question was divided into two sub-questions. The first sub-question aims to highlight the existing approaches, while the second sub-section aims to describe the identified research challenges. In the following, we discuss these sub-questions separately:

*RQ 2.1. What are the proposed different automated user assistance solutions?*
Similar to the categorization of the domains in which user assistance has been applied we have categorized the solution approaches using descriptive and qualitative synthesis approach. This resulted in the following five main categories *Modeling, Process, Framework, Architecture* and *Tool*. Within each category, we further derived the sub-categories. The categories together with the sub-categories, and the primary studies are shown in Table 6.

The category *Modeling* presents the studies on adopted models in user assistance. The category *Process* categorizes the studies that focus on the adopted process in user assistance. The category *Framework* represents the studies that propose or discuss user assistance frameworks. The category *Architecture* presents the studies that discuss the architecture (multi-agent, reference) of user assistance systems. Finally, the category *Tool* discusses the development of tools for user assistance.

Within each category we have further identified sub-categories in which the primary studies have done research and proposed the corresponding solutions. The middle column of Table 6 therefore identifies both the identified research areas and the solution areas.

We observed that each primary study usually applies more than one of these specific categories, meaning that the provision of user assistance is achieved through the integration of several approaches. A further important observation is that most of the primary studies include and employ at least one of the concepts under the category *Modeling*. In fact this is not strange since for defining a process, framework, architecture, or tool very often the underlying models are required. An important domain of modeling that requires attention is agent-based models in which we could identify nine primary studies. It appears that for providing user assistance very often agents are adopted that take care of providing the appropriate help.

Study K deals with the recommendation agents in browsing based on the integration of user profiles, navigational patterns and contextual elements. The main theme here is that pro-active and context-aware retrieval in which relevant documents are automatically presented to users according to their activities is based on the knowledge about active user goals. The study O is grounded on the hypothesis that there are a considerable number of users willing to use agents provided that they know what an agent is all about. The idea here is to develop user assistance software that is highly customizable and adaptable to the user configurations and preferences. They tried to simplify the instruction process of an agent as close as possible to natural language specifications. In the study P, hypothesized intentions of users are the indicators of the accuracy of their workflows. Their graphical user interface mechanism is supposed to intervene undesirable situations and provide automated assistance. The authors emphasize the users' reluctance to use help even in problematic cases. The study F proposes an agent-based framework as a recommendation agent that offers related documents with respect to users' responses in an ad-hoc fashion. The authors' of study I propose a reflective architecture comprised of a set of cooperative agents for modular design of application assistance software. Using several autonomous agents each dedicated to a single task is employed to provide automated user assistance. The study R is based on the adoption of agents in web authoring tasks. Two tools intended to capture the user's preferences and assist him/her throughout the interaction are designed, and as such minimizing the effort in webpage authoring. Unobtrusive and pro-active user assistance is pronounced also in the study BB which is based on a structured pipeline of perception, sensor interpretation, intention analysis, strategy synthesis, and actuation. The study U is of somewhat unlike nature that it is a novel concept for cognitive assistance and training in manual industrial assembly aimed at designing a mobile, personal system, for the purposes of task solving and tool handling. Handling the increased complexity in industrial processes is tried to be solved by, in a sense, an agent-based fashion.

TABLE 6: CATEGORIES OF AUTOMATED USER ASSISTANCE SOLUTIONS

| Main Category | Specific Solution and Research Category | Studies |
|---|---|---|
| Modeling | Agent-based models | F, J, K, M, O, P, R, U, BB |
| | Ontology-based models | D, D, F, R, EE |
| | Goal-based models | F, P, Z, BB |
| | Cognitive ergonomics models | M, U |
| | Interaction modalities | B, S |
| | Domain-specific models | O |
| | User models | X, Q, Y, Z, BB |
| | Behavior models | Q, B, W, Y, K, Z, T, EE |
| | Interest models | W, Y, K, Z, T, Q |
| Process | Case-based reasoning | DD |
| | Goal recognition | P |
| | Human-plausible reasoning | P, H |
| Framework | Dynamic framework | B, CC |
| | Agent-based framework | F |
| | Model-driven framework | O, EE |
| | Probabilistic framework | A |
| | Component-based framework | BB |
| | Conceptual framework | N |
| Architecture | Multi-agent architecture | J, O |
| | Reference architecture | O, EE |
| Tool | Taxonomy-based tools | N |
| | Client-side | L |
| | Mobile tool applications | G, V, S |
| | Framework-based tools | Y, O, EE |
| | Add-on tools | I |

Several primary studies indicate the need for providing ontologies of the subject and the required help to support user assistance. We can also observe that goal-based solutions are partially merged with agent-based proposals as we have two studies in this fashion. The study U is related to manual tasks in industry, and somehow the solution here is both agent-based and cognitive in terms of ergonomics. User model, user behavior models, user interest models, inference components and collaboration components are mostly used in the primary studies that are in the group of collaborative environments.

From the perspective of the individual studies we can observe that the studies Q, T and Y are the most active participants under these specific categories. Study AA uses an ensemble of specific solution methods within a pipelining mechanism such as agent-based, goal-based and user models which will further lead the study to a component-based framework structure. Study O is a somewhat single performer in the fields of domain-specific models, model-driven frameworks and reference architectures which are the concepts of software engineering paradigm. Study P aims at goal-recognition and corresponding human-plausible reasoning.

Eleven of all the specific sub-categories contain only one primary study, indicating the various solution approaches among the primary studies.

From the perspective of tool support we can observe that nine of the primary studies conclude at the design and development of automated tools for the purpose of user assistance. Additionally, almost all of these studies are inspired from modeling approaches.

The given categories could act as a roadmap for the researchers and practitioners in the context of automated user assistance systems. There is a diverse range of solution methods for broad target domains, and employing the discovered methods, even in a multidisciplinary fashion, would introduce fine-granular automated user assistance proposals.

*RQ 2.2. What are the implications of automated user assistance solutions for future research?*
This question is aimed at revealing the implications behind the primary studies for further advancements. As stated before, the identified sub-domains of the solution categories of Table 6 define the solution area as well as the research directions in the domain. In addition to this we have also used the categories identified in RQ1, and presented the implications for each group accordingly in order to further the data synthesis. Hereby, some primary studies appear in more than one domain. In that case we will describe the study in which the primary study is more explicitly described. We discuss each domain separately.

**Education**
In the domain of education a substantial focus has been provided to assisting tasks performed by students or engineers.

The main themes are based on providing adequate help in the cases where the concepts of tasks can be handled as granules that bring fine-grained consideration of task experience. In other words, task experience is grounded on a modeling approach, specifically ontology-based models.

In study B, the authors focus on the multi-modality evidences for automated user assistance such as physical appearance features, physiological measures, user performance and observed behaviors in order to compose a dynamic probabilistic inference model for stress recognition. They imply that a dynamic influence diagram model can successfully recognize human stress and provide automated user assistance seamlessly. Thanks to the just-in time assistance provided, they keep the users in a positive state by holding stress levels down.

In study D, the problem of finding help and the appropriateness of help for a specific object is discussed. The study D puts forward that ontology-based approach should not be bounded to task-experience dimensions in task-specific environments. Here, the authors consider the possibility of sharing ontologies for help systems across applications. Also, the automated identification of right ontologies associated with user's task experience is another future research direction where user models exploit task experience models.

The main theme of study M is to assist teachers by a check mechanism of student participations within a collaborative distance learning environment. For this purpose, an intelligent agent integrated with a web-based distance learning platform is designated. The use of automated user assistance increased the number of collaboration conflicts detected. The means of evidence behind here is that the conflict detection accuracy was validated both with artificial data and with a controlled group of users in a real course. They leave a concrete future work that is the experimentation with larger number of courses and groups of students in order to capture more conflicting cases and tune the model accordingly.

One of the interesting categories of primary studies in education is remote experimentation. The studies AA and DD, where the former repeats the latter by the same authors, focus on intelligent user help in the context of remote experimentation. Study AA presents an intelligent context specific help system in terms of its embryonic stages of architecture and design. Their implication is that the upward tendency of the use of automated user assistance solutions in remote experimentation environments will certainly increase. In the study DD, the proposed method offers considerable benefits in the field of remote experimentation. They initiated the development of an intelligent context-sensitive help system that shows promising results. The facilitation of collaborative experimentation between students is achieved through the addition of remote cooperative working functionality to the remote laboratory. The matter of remote learning from a students' perspective is that context specific help is necessary at some point of experimentation.

**Control Systems**
In the domain of control systems we can identify the effort for developing tool support for user assistance. Hereby, the need for providing help based on the state of the process is an interesting direction.

The study BB deals with the core question of providing automated user assistance in smart environments. The authors suggest the integration of modeling and simulation efforts for the different types of models at different levels of abstractions, whereas they exploited them separately. Also, a component-based modeling and simulation framework for smart environments would be the major implication of this study for future research.

Study EE describes proposes an aspect-oriented tool framework that can be used to develop process-sensitive embedded user assistance for multiple applications. The framework provides tools for defining the process model, defining guidance related to process steps, and modularizing and weaving help concerns in the target application for which user guidance needs to be provided.

**Manufacturing**

In this domain the main idea was to apply user assistance and less focus has been provided in enhancing technical concepts of user assistance.

The authors of study U highlight the lack of methods helping operators in executing complex, manual assembly tasks, and an overview about personal cognitive assistance in production environments. To the best of their knowledge, in this field, the only available instruments are training courses, text-based documentations and learning-by-doing, which are time-consuming, extensive and inadequate. The main implication is that further advancements are to be in favor of humans instead of substituting them.

Study M in this category treats user assistance as a somewhat fuzzy concept, and this concept is said to be requiring derivations from cognitive ergonomics. This study is a framework level of proposal leading to a comprehensive taxonomy.

**Collaborative Design**

The studies in the collaborative design category have focused on either collaborative design or collaborative learning environments, and they deal with the appropriateness of automation of user assistance in these environments. Collectively discussing the main themes stated under this category, intelligent user assistance and mediums like software agents have been recognized as a promising approach to implement collaborative systems. In collaborative environments, using cognitive user models, especially user interest and user behavior models, is proposed along with the utilization of inference, knowledge update and collaboration components. The models and components are the basis of personal assistant agents from which we can derive flexibility and adaptability to effectively work with the corresponding users to achieve their goals in goal-directed collaborative tasks. In other words, the main idea here is to create user-adaptive environments within collaborative systems.

The authors of study Q state that the design of personal assistant agents in collaborative environments is a new research area. The major issues to be taken into account are user modeling, reasoning and design making, and collaboration mechanisms. They specifically focused on encapsulating the specific collaboration mechanisms into a generic collaboration component and as such providing automated user assistance in collaborative design environments. The proposed approach is said to be effective for the provision of automated user assistance. Also in the study Y, the same authors propose an agent-architecture in which the collaboration between the agents helps the engineers to finish the collaborative design task successfully. The study Z is based on exploiting the user model to capture the user's interests and behaviors and as such providing automated user assistance. According to the authors, in real world engineering design environments, the proposed approach, which is based on a collaborative personal assistant agent framework, would be used. They also suggest an approach in which the users are able to customize their user models in a collaborative design environment.

The study O reveals a software engineering issue related to agent-based environments which is the user customizations issue. The findings indicate that the choice of the dimension in which the software architecture will be modularized is of utmost importance. From this perspective, the authors bring out the need for better software architectures, where security and privacy issues are also considered, to build personalized user agents.

The study T states a long-term goal and a possible implication for research that is to develop a general collaborative personal assistant agent framework applicable to various collaborative engineering environments.

**Healthcare**

Within the domain of healthcare we can identify only two studies among the selected primary studies. These focus on providing decision and mathematical models for supporting user assistance.

**Mobile Applications**

In the category of mobile devices, we have three of the primary studies. Study F is related to the declarative description of actions, tasks, and solution methods, by which hybrid planning allows for the generation of knowledge-rich plans of action. First, the authors show a goal-based approach, which is based on automated reasoning techniques, on user plans, just like in plan recognition, and a possible speech dialog. Their major conclusion is that assistance in this method is provided only in case it is needed, and as a future work, a tutoring system such as a proactive electronic instruction manual will be useful for the purpose of automated user assistance.

Study V shows an approach which exploits patterns in mobile usage. The authors applied this mechanism as an automated user assistance solution, and they propose it as a highly-used and interesting component to the mobile phones.

Study S is based on the provision of automated user assistance for mobile devices in ubiquitous computing environments. The proposed approach handles the group of inexperienced users since their acquaintance improves slowly. A possible advancement would be the evaluation of this approach based on some user studies along with a broader range of user groups.

**Web**

In this category user assistance seems to have focused on web search and information retrieval. The articles in information retrieval systems got very high scores in terms of the quality assessment. In the context of information retrieval systems automated assistance is defined as a temporal, goal-driven dialogue of expressions, actions or responses. The studies K and H are based on determining whether automated user assistance improves searching performance. They are founded upon the general opinion that there is a lack of empirical evidence about the instrumentality of automated assistance during the search process. The time users need assistance and the type of assistance they look for are the factors under consideration. For example, it is stated that users seldom make use of advanced search features without even knowing how to exploit them. The extent and the circumstances to which automated assistance is of actual benefit in information searching process is indeterminate.

The study L in the field of information retrieval proposes the design of a general-purpose automated assistance application using implicit feedback. The noteworthy finding here is that automated assistance systems may improve the Web searching performance by offering more number of relevant documents. For future studies, predictability is brought forward as a concern to be considered in these systems since the users utilize these systems in a predictable manner. Also, instead of offering automated user assistance at the query level, the session level assistance along with a more personalized and targeted approach is practicable being more advantageous. The main implication of the study I is that detecting the patterns of user–system interaction, we can customize automated user assistance systems providing just-in-time guidance.

User assistance for web applications seems to be also very often discussed using agent-based approaches. In the domain of agent-based environments, there are nine studies that are collectively assessed in terms of the implications. In the study J, the authors aim at ensuring coordination among various assistants working for the same application by offering a great degree of flexibility and modularity in the design and implementation of assistants. They show the appropriateness of an assistant set based on reflective software architecture for a web browser which will support e-commerce activities. They also discussed the modularity and reusability issues of their proposed assistants. The run-time integration of highly-cohesive assistants and the customization of them for a variety of applications are the major findings.

It is stated in the study K that the use of browsing assistants enables us to capture interests from an ongoing user activity and to detect out-of-context interests. The proposed approach is to specify and associate browsing activities in user profiles. In this way, consistent, comprehensive and meaningful contexts can be identified. Also, the empirical results in this study indicate that the extraction of association rules describing browsing patterns at a conceptual level assists in estimating user interests in a browsing session.

The study F proposes the approach of recommending related documents according to the user feedback inferred from similar-page searching mode. They highlight an implication that capturing semantic relations through some semantic-based improvements in the architecture and defining relative ontology among concepts during Web navigation are the open research directions.

**User Interface Design**

In the category of adaptive, multi-layer and multi-dimensional user interfaces the four studies adopt a multidisciplinary approach.

Study C is more like a survey of modularity dimensions of online help, the authors discuss the variation across multiple levels of user experience. They specified why-what-how dimensions, and the use of these dimensions in the task-application-user dimension is an open research challenge. Also, the authors suggest a rating mechanism within an automated user assistance system in which the users evaluate help content, identifying correctness of help with some reasons of errors. Also, dynamic and user-settable levels could enhance user experience in which users customize the level of explanation through slider bars in a multi-layer, multidimensional interface. This mechanism is also an implication for future research.

Study W brings out episode-based learning compared to other approaches for user assistance. Without a definite conclusion based on a large sample of subjects, their method presents automated learning and personalized adaptation according to some empirical results. Their implication for future research is the extension of this approach as a generic tool in an application- independent context for reusability.

In study X, the research scope is limited to Artificial Intelligence, User Modeling and Human-Computer Interaction. The authors suggest multi-disciplinary approaches in the development of intelligent user interfaces, and they emphasize an open research challenge that is the limited amount of empirical evaluation of adaptive systems. In order to manifest the advantages of adaptive interfaces compared to non-intelligent interfaces, more research should be carried out together with strong empirical evidences. The study E focuses on the concept of semantically transparent interfaces to provide automated user assistance based on ontology-structured text collections. They imply the possibility of integrating semantic transparency into multi-layered interfaces. The future advancements will certainly fall into the intersection of Human-Computer Interaction and Knowledge Management.

Study R provide a particular solution which tries to find out the boundaries of WYSIWYG (What You See Is What You Get) approach, rather than a universal one to the issue of end-user authoring, but. The approach is multidisciplinary in a way that it is based on Programming by Example and Model-Based User Interfaces paradigms.

Study P states that the incorporation of intelligence to a graphical user interface will bring out much more productive users. In order to adapt this approach to a different domain, the kinds of errors that users usually make should be revealed by an empirical study by which the reasoning mechanisms are tuned accordingly. It is empirically assessed that the users of an interface may not be aware of the situations in which they actually need help.

Study D, semantic transparency concept is introduced as a user interface property. In study W, semi-autonomous manipulation of the software systems by adaptive user interfaces is proposed. The main motivation behind this is that the more intuitively a user interface is designed, the more effectively users operate a software system.

In the study CC, manual control of an interactive system is achieved through a predictive haptic user assistance method. It is related to offering real-time guidance for the situations such as animation control and driving

## 5. DISCUSSION

The main threats to validity (Dyba & Dingsoyr, 2008; B. Kitchenham & Charters, 2007; MacDonell, Shepperd, Kitchenham, & Mendes, 2010) of this review are publication and selection bias, and data extraction and classification.

The publication bias indicates the case in which researchers are more likely to publish positive results and refrain from publishing studies that have negative results. To cope with this publication bias Kitchenham et al. (B. Kitchenham & Charters, 2007) recommend to search also company journals, grey literature, conference proceedings and the internet. We have applied this approach which indeed led us to new papers that we could not identify in our regular search. We used a quasi-gold standard (QGS) to form an optimal search strategy. We applied word frequency and statistical analysis tools on a well-controlled and piloted set of studies in order to capture better keywords for the review. However, it should be noted that there were some other keywords being discipline, or category-specific in our case. We performed the inclusion/exclusion procedures on a well-established screening of primary studies. We included both qualitative and quantitative studies in almost all respects. Also, we translated the results of primary studies in each group as much as possible in order to attain uniformity at least in specific categories. As such, we tried to reduce the impact of the publication bias as much as possible by adopting the guidelines and criteria as defined in the studies on systematic literature reviews.

The inclusion and exclusion criteria are selected by the researchers who perform the systematic literature review. A subjective approach towards defining the selection criteria and selecting the primary studies for further consideration, can introduce a threat to validity in this study. For reducing the bias with respect to the definition of the selection criteria we use the quasi-gold standard approach as defined by Zang et al. (Zhang et al., 2011). Hereby, we first picked a random set of 10 studies and each of the researchers defined the selection criteria. These criteria were validated together and the final set of exclusion/inclusion criteria was defined.

For reducing the selection bias for selecting the primary studies, the evaluation and the selection of the primary studies were performed separately by three researchers. Each researcher recorded also the reasons of acceptance or rejection for all the considered studies. Later on the evaluated list of primary studies of each researcher was compared with that of the other researchers. In case of differences we discussed the paper in detail and came with the final decision.

After the primary studies have been evaluated and selected the relevant data must be extracted for deriving the review results. Hereby defining the data extraction criteria and classification model is very important. To define the data extraction model we first read a set of randomly selected primary study papers. Each of use defined an initial data model based on the research questions that we had defined. Later on, we compared the different data extraction models, discussed the differences and decided on the data extraction model. After that we applied the data extraction model to a set of primary studies and checked whether we could derive the answers to the research questions with the adopted data extraction model. We applied this several times and after a number of iterations and discussions we decided on the final data extraction model.

## 6. CONCLUSION

A relatively broad interest can be observed for automated user assistance in a diverse range of domains. To the best of our knowledge, no previous systematic literature study has been performed before for the domain of automated user assistance systems. We tried to reveal the body of multidisciplinary research on this field by systematically analyzing the published literature since 2002. We reviewed 575 papers that are discovered using a well-planned review protocol, and 31 of them were assessed as primary studies related to our research questions.

Considering the diversity of primary studies with respect to main themes, they were categorized into groups according to the nature of each research question. We have reported the benefits of automated user assistance solutions in several domains along with the implications for the future research and practice. The strength of evidence appeared to be relatively good in spite of the diversity of both target domains and solution methods.

This systematic review is originally based on the findings of individual primary studies that address the introduction and adoption of automated user assistance techniques. The majority of the papers present positive statements in support of their findings towards automated user assistance. The qualitative data reported is mostly in parallel with the quantitative results.

The benefits of automated user assistance are strongly emphasized in software intensive systems in which the interaction with the users is of vital importance. The main argument here is that we can ensure better user experience through better user assistance. The development and need of autonomous agents for user assistance is increasingly growing, especially in the industry. Context awareness and intelligent user interfaces provide the users just-in-time assistance by which they stick with the concerned workflows.

A further significant finding of this review is that there are still many open research issues that need closer attention. Our work can be considered as a roadmap to identify the current state-of automated user assistance and as such pave the way for further research in this domain.

**APPENDIX-I LIST OF PRIMARY STUDIES**

**APPENDIX-II STUDY QUALITY ASSESSMENT**

| | | Reporting | | | Relevance | Rigor | | | Credibility | | |
|---|---|---|---|---|---|---|---|---|---|---|---|
| | Aim | Scope, Context and Design | Evaluation Rationale | Description of study participants | Implications in practice and research | Validity and reliability of variables | Explicitness of measures | Adequacy of reporting | Creditability, validity and reliability | Limitations | |
| Primary Study | Q1 | Q2 | Q3 | Q4 | Q5 | Q6 | Q7 | Q8 | Q9 | Q10 | TOTAL (out of 10) |
| A | 1 | 0.5 | 1 | 1 | 0.5 | 1 | 0.5 | 1 | 0.5 | 0.5 | 7.5 |
| B | 1 | 1 | 1 | 0.5 | 0.5 | 0.5 | 1 | 1 | 1 | 0.5 | 8 |
| C | 1 | 1 | 0.5 | 1 | 1 | 0.5 | 0 | 0 | 0.5 | 0 | 5.5 |
| D | 1 | 1 | 1 | 0.5 | 1 | 0.5 | 0.5 | 0.5 | 1 | 0 | 7 |
| E | 1 | 1 | 1 | 0.5 | 1 | 0.5 | 0.5 | 0.5 | 0.5 | 0.5 | 7 |
| F | 1 | 1 | 1 | 0.5 | 0.5 | 0.5 | 1 | 1 | 0.5 | 0 | 7 |
| G | 1 | 1 | 0.5 | 1 | 0.5 | 0.5 | 0.5 | 1 | 0.5 | 0.5 | 7 |
| H | 1 | 1 | 1 | 0.5 | 1 | 0.5 | 1 | 1 | 0.5 | 0 | 7.5 |
| I | 1 | 1 | 1 | 1 | 0.5 | 1 | 0.5 | 1 | 0.5 | 0.5 | 8 |
| J | 1 | 1 | 0.5 | 0.5 | 1 | 0 | 0.5 | 1 | 1 | 0.5 | 7 |
| K | 1 | 1 | 1 | 1 | 1 | 1 | 1 | 1 | 0.5 | 0 | 8.5 |
| L | 1 | 1 | 1 | 1 | 1 | 0.5 | 0.5 | 1 | 1 | 0 | 8 |
| M | 1 | 1 | 1 | 0.5 | 1 | 1 | 0.5 | 1 | 0.5 | 0 | 7.5 |
| N | 1 | 0.5 | 0.5 | 1 | 1 | 0 | 1 | 0.5 | 0.5 | 0 | 6 |
| O | 1 | 1 | 1 | 1 | 1 | 1 | 0.5 | 0.5 | 0 | 0.5 | 7.5 |
| P | 1 | 1 | 1 | 1 | 1 | 0.5 | 0.5 | 1 | 1 | 0.5 | 8.5 |
| Q | 1 | 0.5 | 0.5 | 1 | 0.5 | 0.5 | 0.5 | 0.5 | 0.5 | 0 | 5.5 |
| R | 1 | 1 | 1 | 0.5 | 1 | 1 | 0.5 | 1 | 1 | 0.5 | 8.5 |
| S | 1 | 1 | 1 | 0.5 | 0.5 | 0.5 | 0.5 | 0.5 | 0 | 0.5 | 6 |
| T | 1 | 1 | 1 | 1 | 0.5 | 0.5 | 0.5 | 0.5 | 0.5 | 0 | 6.5 |
| U | 1 | 0.5 | 0 | 1 | 0.5 | 0 | 0.5 | 0 | 0.5 | 0 | 4 |
| V | 1 | 0.5 | 0.5 | 0.5 | 1 | 0.5 | 0 | 0.5 | 0.5 | 0 | 5 |
| W | 1 | 0.5 | 1 | 1 | 0.5 | 0.5 | 1 | 0.5 | 0 | 0.5 | 6.5 |
| X | 1 | 1 | 1 | 0.5 | 1 | 0 | 0.5 | 0.5 | 1 | 0.5 | 7 |
| Y | 1 | 1 | 1 | 0.5 | 1 | 0 | 0.5 | 0.5 | 0.5 | 0.5 | 6.5 |
| Z | 1 | 1 | 0.5 | 0.5 | 1 | 1 | 1 | 1 | 0.5 | 0.5 | 8 |
| AA | 1 | 1 | 1 | 1 | 1 | 0.5 | 0.5 | 0.5 | 0.5 | 0.5 | 7.5 |
| BB | 1 | 1 | 1 | 0.5 | 1 | 0.5 | 0.5 | 0.5 | 0.5 | 0.5 | 7 |
| CC | 1 | 1 | 0.5 | 1 | 0.5 | 0.5 | 0.5 | 0.5 | 0.5 | 0.5 | 6.5 |
| DD | 1 | 0.5 | 0 | 1 | 0.5 | 0 | 0.5 | 0.5 | 0.5 | 0 | 4.5 |
| EE | 1 | 1 | 1 | 0.5 | 1 | 0.5 | 1 | 1 | 0.5 | 0.5 | 8 |

**APPENDIX-III SEARCH STRINGS**

| Repository | String |
|---|---|
| ACM | *Title*:("user" AND "assistance") AND ("context sensitive" OR "context-sensitive" OR " process-sensitive" OR "process sensitive" OR "context aware" OR "context-aware" OR "embedded" OR "intelligent" OR "adaptive")<br>*Abstract*:("user" AND "assistance") AND ("context sensitive" OR "context-sensitive" OR " process-sensitive" OR "process sensitive" OR "context aware" OR "context-aware" OR "embedded" OR "intelligent" OR "adaptive")<br>*Keywords*:("user" AND "assistance") AND ("context sensitive" OR "context-sensitive" OR " process-sensitive" OR "process sensitive" OR "context aware" OR "context-aware" OR "embedded" OR "intelligent" OR "adaptive") |
| IEEE | ("user" AND "assistance") AND ("context sensitive" OR "context-sensitive" OR " process-sensitive" OR "process sensitive" OR "context aware" OR "context-aware" OR "embedded" OR "intelligent" OR "adaptive") |
| ISI Web of Kn. | TS=(("user assistance") AND ("automated" OR "context sensitive" OR "context-sensitive" OR " process-sensitive" OR "process sensitive" OR "context aware" OR "context-aware" OR "embedded" OR "intelligent" OR "adaptive")) |
| Science Direct | Abstract, Title, Keywords:<br>("user" AND "assistance") AND ("context sensitive" OR "context-sensitive" OR " process-sensitive" OR "process sensitive" OR "context aware" OR "context-aware" OR "embedded" OR "intelligent" OR "adaptive") |
| Springer | Title, Abstract:<br>("user" AND "assistance") AND ("context sensitive" OR "context-sensitive" OR " process-sensitive" OR "process sensitive" OR "context aware" OR "context-aware" OR "embedded" OR "intelligent" OR "adaptive") |
| Wiley Interscience | Publication Titles, Article Titles, Abstract, Keywords:<br>("user" AND "assistance") AND ("context sensitive" OR "context-sensitive" OR " process-sensitive" OR "process sensitive" OR "context aware" OR "context-aware" OR "embedded" OR "intelligent" OR "adaptive") |

**APPENDIX-IV DATA EXTRACTION FORM**

| Study description | Extraction element | Contents |
|---|---|---|
| General Information | | |
| 1 | ID | Unique id for the study |
| 2 | SLR Category | ○ Include  ○ Exclude |
| 3 | Title | Full title of the article |
| 4 | Date of Extraction | The date it is added into repository |
| 5 | Year | The publication year |
| 6 | Authors | |
| 7 | Repository | ACM, IEEE, ISI Web of Knowledge, Science Direct, Springer, Wiley Interscience |
| 8 | Type | ○ Journal ○ Conference ○ Other (dissertation, grey literature etc.) |
| 9 | Does it repeat already reviewed paper? | ○ Yes (Repeated ID)  ○ No |
| Relevance | | |
| 10 | Does it relate to a specific field of computer science? | ○ Yes  ○ No ○ To some extent |
| 11 | Does it relate to user assistance? | ○ Yes  ○ No ○ To some extent |
| 12 | Subjects | ○ Academics ○ Industry/Real world |
| Study Description | | |
| 10 | Main theme of the study | |
| 11 | Motivation for the main theme | |
| 12 | Study aims | |
| 13 | Targeted domain | Task-specific environments, Adaptive, multi-layer and multi-dimensional user interfaces, Collaborative Environments, Interactive systems, Information Retrieval Systems, Agent-based environments, Mobile devices, Learning Environments for Remote Experimentation |
| 14 | Automated user assistance solution used | |
| 15 | Examples of application of solution | Critical, Web-based, Portable appliances, Non-functional concerns, Real-world, Computer-aided |
| 16 | Research method used | Case study, Multiple-case study, Experiment, Benchmarking, , Survey |
| 17 | Assessment approach | Qualitative, Quantitative or Both |
| 18 | Findings | |
| 19 | Constraints/limitations | |
| 20 | Implications for future research | |
| 21 | Major conclusions | |
| Evaluation | | |
| 22 | Personal note | The opinions of the reviewer about the study |
| 23 | Additional note | Publication details (supported by grants etc.) |
| 24 | Quality Assessment | Detailed quality scores |